%% file: rev.tex
\documentclass[prb,aps,amssymb,showpacs,twocolumn]{revtex4}
\usepackage{amsmath}
\usepackage{amssymb}
\usepackage{amsthm}
\usepackage{amsfonts}
\usepackage{algorithmic}
\usepackage{enumerate}
\usepackage{latexsym}
\usepackage[dvips]{graphicx}

\newcommand{\beq}{\begin{equation}}
\newcommand{\eneq}{\end{equation}}

\input{epsf}

\begin{document}

\tolerance 10000

%\draft

\title{Nearly Insulating Strongly Correlated Systems: Gossamer Superconductors 
and Metals}

\author { Bogdan A. Bernevig$^\dagger$, George Chapline$^\square$, Robert B. 
Laughlin$^\dagger$, Zaira Nazario$^\dagger$ and 
David I. Santiago$^{\dagger, \star}$ }

\affiliation{ $\dagger$ Department of Physics, Stanford
         University,
         Stanford, California 94305 \\ $\square$ Physics and Advanced 
         Technologies Directorate\\ 
         Lawrence Livermore National Laboratory, Livermore, CA 94550.\\
         $\star$ Gravity Probe B Relativity, Stanford University Mission, 
         Stanford, California 94305}
\begin{abstract}
%\vspace*{-1.0truecm}
\begin{center}

\parbox{14cm}{Recently a new phenomenological Hamiltonian was introduced to 
describe the superconducting cuprates in which correlations and
on-site Coulomb repulsion are introduced by partial Gutzwiller projection. 
This Gossamer Hamiltonian has an exact ground state and differs from 
the t-J and Hubbard Hamiltonians in possessing a powerful attractive 
interaction among electrons responsible for Cooper pairing in a d-wave channel.
It is a faithful description for a superconductor with strong on-site  
electronic repulsion. The supercodnucting tunnelling gap remains intact and 
despite on-site repulsion. Near half-filling the Gossamer superconductor 
with strong repulsion has suppresed photoemission intensities and superfluid 
density,  is unstable toward an antiferromagnetic insulator and posseses an 
incipient Mott-Hubbard gap. The Gossamer technique can be applied to metallic
ground states thus possibly serving as an apt description of strongly 
correlated metals. Such a Gossamer metallic phase, just as the Gossamer 
superconducting one, becomes arbitrarily hard to differentiate from an 
insulator as one turns the Coulomb correlations up near half-filling. 
Both the metallic and superconducting states undergo a quantum phase
transition to an antiferromagnetic insulator as one increases the on-site 
Coulomb repulsion. In the Gossamer model we reach the critical point
at half-filling by fully projecting double occupancy. Such a critical point
might be the Anderson Resonating Valence bond state.}

\end{center}
\end{abstract}


\maketitle

\section{Introduction}

The high temperature superconducting cuprates are unusual in large part because
they develop from  correlated antiferromagnetic insulators when doped. 
The antiferromagnetism and insulation are caused by the strong on-site Coulomb 
repulsion among the Copper d-electrons. Although these electron correlations 
have been postulated to be the key ingredient to the superconductivity in the 
cuprates\cite{rvb}, Laughlin\cite{bob} and some of us\cite{us} have recently 
suggested that correlation effects compete and are detrimental 
to the superconductivity. A specific model\cite{bob,us} was proposed,
the Gossamer superconductor, in which the spectral function will evolve with 
decreasing doping toward that of an insulator with two Hubbard bands, a 
Hubbard gap and an ever fainter band of mid-gap states\cite{fuji} 
corresponding to the collapsing superfluid density. Despite this
behavior the model has an exact d-wave superconducting ground state 
{\it for all dopings} up to the half-filled undoped state. This is achieved 
through partial Gutzwiller projection of the superconductor.

Underdoped cuprate superconductors have superfluid density and transition 
temperature that are proportional to each other and vanish 
linearly\cite{uemura} with doping. The proportionality is consistent with the 
transition being an order-parameter phase instability\cite{emery}: the 
superconducting transition temperature, T$_c$, is lower than the pairing 
temperature because the superconductor is becoming increasingly unstable 
to loss of phase coherence due to the small superfluid 
density\cite{emery,dynes}. We thus expect the measured pseudogap\cite{timusk} 
in underdoped cuprates to be caused by preexisting Cooper pair correlations. 
This idea is supported by optical sum rule studies\cite{uchida}, the giant 
proximity\cite{drew}, and the recent heat-transport measurements\cite{ong} 
showing superconducting vortex-like effects above the transition temperature.

Although the motivation for the Gossamer model is the superconducting cuprates
for which it provides a very apt description, the behavior should be
general to all superconductors with strong on-site repulsion and even to 
strongly correlated metals\cite{kollar,us2}. Such strongly correlated
superconductors will be {\it spectroscopically identical} to doped 
Mott-insulators close to half-filling, except for a small amount of conducting 
fluid corresponding to the dephased superconductor. This naturally accommodates
experiments that hint at conduction in the supposedly antiferromagnetic
insulating phase\cite{ando0} with carrier density proportional to doping, 
the existence of a d-wave node deep in the underdoped regime\cite{fuji} 
detached from the lower Hubbard band and simply materializing at mid-gap with
increasing doping is increased from zero, and a d-wave gap in the quasiparticle
spectrum that grows monotonically as the doping decreases and saturates at a 
value of about 0.3 eV \cite{marshal, ding, rbl, timusk}.  

The idea that the ``insulator'' might actually be a tenuos
superconductor is implicit in the early ideas of the Anderson on the resonating
valence bond (RVB) state\cite{rvb,gros,shiba,zhang,zg} and in more recent 
work\cite{rand,zg2,zg3}. This idea has always run counter to the basic 
premise of RVB theory that superconductivity should be a universal aspect of 
quantum antiferromagnetism. The conventional spin density wave ground state, 
which contains no superconductivity, is a faithful prototype for a 
quantum antiferromagnet contradicting this basic premise. Thus not all 
antiferromagnets are superconductors but some of them are and they constitute a
separate class of antiferromagnets: they contain a strong attractive 
interaction giving rise to superconductivity and a tiny background superfluid 
density due to strong electron correlations. The Gossamer Hamiltonian 
stabilizes superconductivity in the latter antiferromagnets over the 
spin density wave state. This suggests that Coulomb interactions are not 
sufficient to explain cuprate superconductivity.

Finally, we will discuss the application of the Gossamer technique to
correlated metallic systems\cite{kollar,us2}. The phase diagram of Vanadium 
sesquioxide, V$_2$O$_3$, and other Mott insulators is such that the 
antiferromagnetism and insulation disappear with pressure at $T=0$\cite{vanad} 
and/or doping. With increasesing temperature, the antiferromagnetic order 
melts and one is left with anisulating-like spin disordered phase. Applying 
pressure to this insulating-like phase it undergoes a first order phase 
transition into a metal. The line of first order transitions between the metal
and the insulator-like phase terminates at a critical point $(T_c,
P_c)$ as it happens in f-electron compounds\cite{ce}. This suggests that the 
metal and the insulating-like phase cannot be fundamentally different as one 
can go continuously from one into the other above the critical temperature. We 
thus propose that the spin disordered insulator-like phase is a bad tenuous 
``Gossamer'' metal rather than a true insulator. 

The Gossamer metal will have properties similar to the Gossamer superconductor.
The number of conducting electron states and with it, the density of states at 
the Fermi level collapses to zero near half-filling when the correlations are 
strong. The missing  spectral weight forms Mott-Hubbard bands.  
Such a ``Gossamer'' metal cannot be differentiated from an insulator 
except at the lowest possible temperatures in which, at
least for the case of V$_2$O$_3$, antiferromagnetic order is stablished with 
the material becoming truly insulating. If antiferromagnetism were not to
intervene, we predict the resistivity to saturate to a large
but finite value at zero temperature absent localization effects.
We propose that this is happening for the disordered insulating-like phase of  
correlated electron systems\cite{us2}.

\section{Gossamer Superconductor}

The Gossamer superconductor is defined as a superconducting
ground state which contains Coulomb correlations introduced by
a partial Gutzwiller projection which suppresses the probability of having
two electrons on the same site:

\begin{equation}
\Pi_{\alpha} = \prod_j z^{(n_{j \uparrow} + n_{j \downarrow}) /
2}_{0}(1 - \alpha_0 n_{j \uparrow} n_{j \downarrow}) \; \; \; .
\end{equation}
\noindent $ 0 \le \alpha_0 < 1 $ is a measure of how effective the
``projector'' hinders double occupancy and in a real material it will be 
related to Coulomb repulsion. The quantum fugacity, $z_0$, in
the projector is the extra probability of having an electron at
site $j$ after projecting and is necessary in order to keep the
total number of particles constant at $(1-\delta)N$ as one varies
$\alpha_0$, where $\delta$ is the doping. The charge states of a site are
statistically independent and characterized by a fugacity $z$.  The
condition that the total charge on the site be $1 - \delta$ is 

\beq
\frac{2z + 2  z^2}{1 + 2 z +  z^2} = 1 - \delta
\eneq

\noindent before projecting. After projecting the condition becomes 

\beq 
\frac{2z z_0 + 2 (1 - \alpha) (z z_0)^2}{1 + 2 z z_0 + (1 -
\alpha ) (z z_0)^2} = 1 - \delta
\eneq
where $1 - \alpha = (1 - \alpha_0)^2$, giving

\begin{equation}
z = \frac{\sqrt{1 - \alpha ( 1 - \delta^2)} - \delta}
{(1 - \alpha) ( 1 + \delta)} = ( \frac{ 1 - \delta}{1 + \delta} )
\; z_0 \; \; \; .
\end{equation}

\noindent
The parameter $z_0$ is the factor by which $z$ exceeds $(1 - \delta)/(1 +
\delta)$, its value for $\alpha_0 = 0$:
\beq
z_0 = \frac{\sqrt{1 - \alpha(1-
\delta^2)} - \delta}{(1-\alpha)(1-\delta)} \; .
\eneq

The Gossamer superconducting ground state is  

\beq
| \Psi  \rangle =  \Pi_\alpha \; | \Phi  \rangle \label{gs}
\eneq
 
\noindent where $| \Phi \rangle$ is the BCS ground state:

\begin{equation}
|\Phi \rangle = \prod_{\vec{k}} (u_{\vec{k}} + v_{\vec{k}}
c^{\dagger}_{\vec{k} \uparrow} c^{\dagger}_{-\vec{k} \downarrow})
|0> \; \; \; .
\end{equation}

\noindent where $u_{\vec{k}}$, $v_{\vec{k}}$ are the BCS
coheremce factors given by 

\begin{displaymath}
u_{\vec{k}}=\sqrt{\frac{E_{\vec{k}} + 
\epsilon_{\vec{k}}- \mu}{2 E_{\vec{k}}}} 
\end{displaymath}
\beq
v_{\vec{k}}=\sqrt{\frac{E_{\vec{k}}
- (\epsilon_{\vec{k}}- \mu)}{ 2E_{\vec{k}}}}
\eneq

\noindent with dispersion 

\beq
E_{\vec{k}}=\sqrt{(\epsilon_{\vec{k}} - \mu)^2+ \Delta^{2}_{\vec{k}}}
\eneq

\noindent providing the energy of quasiparticle excitations. Here 

\beq
\epsilon_{\vec{k}}= 2t (\cos(k_x a) + 
\cos(k_y a))
\eneq

\noindent is the kinetic energy for a square lattice with spacing $a$, 
$\mu$ is the chemical potential and 

\beq
\Delta_{\vec{k}}= \Delta_o ( \cos (k_x a) - \cos(k_ya))
\eneq

\noindent is the d-wave
superconducting gap as measured for the superconducting cuprates\cite{timusk}. 
In superconductors the coherence factors $u_{\vec{k}}$ and
$v_{\vec{k}}$ are normalized, $u_{\vec k}^2 + v_{\vec k}^2 = 1$, and related 
to the number of carriers  in order to
set the value of the chemical potential. For doped cuprates with doping 
$\delta$, we have 

\begin{equation}
\frac{1}{N} \sum_{\vec k} v_{\vec k}^2 = 1 - 
\frac{1}{N} \sum_{\vec k} u_{\vec k}^2 = \frac{1 - \delta}{2} \label{coh}
\; \; \; .
\end{equation}

For the Gossamer ground state (\ref{gs}) we stay away from full projection 
($\alpha_0 < 1$ always) in order for the partial projector
to have an inverse:
\begin{equation}
\Pi^{-1}_{\alpha} = \prod_j z^{-(n_{j \uparrow} + n_{j
\downarrow}) / 2}_{0}(1 + \beta_0 n_{j \uparrow} n_{j \downarrow})
\; \; \; ,
\end{equation}
\noindent with $\beta_0 = \alpha_0 / (1 - \alpha_0)$. This invertibility
enables us to define the Hamiltonian:

\begin{equation}
{\cal H} = \sum_{\vec{k} \sigma} E_{\vec{k}} B_{\vec{k}
\sigma}^\dagger B_{\vec{k} \sigma}, \;\; \; \; B_{\vec{k} \sigma}
|\Psi \rangle =0 . \label{gossham}
\end{equation}

\noindent  where  

\begin{displaymath}
B_{\vec{k} \uparrow \{\downarrow\}} = \Pi_\alpha b_{\vec{k}
\uparrow \{\downarrow\}} \Pi_\alpha^{-1} = \frac{1}{\sqrt{N}}
\sum_j^N e^{i \vec{k} \cdot \vec{r}_j}
\end{displaymath}

\begin{equation}
\times \biggl[ z_0^{-1/2} u_{\vec{k}} (1 + \beta_0 n_{j \downarrow
\{\uparrow\}} ) c_{j \uparrow \{\downarrow\}} \pm z_0^{1/2}
v_{\vec{k}} (1 - \alpha_0 n_{j \uparrow \{\downarrow \}} ) c_{j
\downarrow \{\uparrow\}}^\dagger \biggr] 
\end{equation}
\noindent with 

\beq
b_{\vec{k}\uparrow \{\downarrow\}} =  u_{\vec{k}}
c_{j \uparrow \{\downarrow\}} \pm v_{\vec{k}} c_{j \downarrow
\{\uparrow\}}^\dagger\eneq 

\noindent the Bogoliubov quasiparticle operators.

The Gossamer ground state (\ref{gs}) is an
exact lowest energy eigenstate of the Gossamer Hamiltonian (\ref{gossham})
by virtue of its
zero eigenvalue and the positivity of the Hamiltonian:

\begin{equation}
\langle \chi | {\cal H} | \chi \rangle = \sum_{{\vec k} \sigma} E_{\vec k}
\langle B_{{\vec k} \sigma} \chi |
B_{{\vec k} \sigma} \chi \rangle \; \geq \; 0
\; \; \; 
\end{equation}

\noindent for any wavefunction $| \chi \rangle$. The Gossamer ground state is 
adiabatically continuable to
the BCS ground state by continously decreasing $\alpha_0$ to zero. Since it 
does not cross a phase boundary in the process, its uniqueness follows from 
the uniqueness of the BCS ground state up to a phase. Therefore the Gossamer 
superconductor describes the same phase of matter as the BCS superconductor. 
The Gossamer superconductor ground state and its low energy excitations map to 
the the ground state and low-lying excitations of a BCS superconductors in a 
one to one manner. 

\section{Quasiparticle Dispersion and Spectroscopy of the Gossamer 
Superconductor}

Let us now consider the quasiparticle excitations of this superconductor.
Since the operators $B_{{\vec k} \sigma}$ no longer have fermionic 
anticommutation relations with their hermitian adjoints, they cannot be used 
to create eigenstates of the Hamiltonian. 
The physical meaning of this is that the quasiparticles
interact. We will approximate the quasiparticle-like eigenstates
with the variational wavefunctions

\beq
| {\vec k} \sigma \rangle = \Pi_\alpha b_{{\vec k} \sigma}^\dagger
| \Phi \rangle \; .
\eneq

\noindent The quasiparticle energy is apporixamate by the expected value

\begin{equation}
\frac{\langle {\vec k} \sigma | {\cal H} | {\vec k} \sigma \rangle}
{ \langle {\vec k} \sigma | {\vec k} \sigma \rangle}
= E_{\vec k} \; \frac{ \langle \Psi | \Psi \rangle}
{\langle {\vec k} \sigma | {\vec k} \sigma  \rangle}
\simeq E_{\vec k} \; .
\end{equation}

\noindent This is an almost miraculous result, the dispersion is unchanged from
the dispersion of the unprojected superconductor. This follows by evaluating 
the relevant norms, which can be done by hand only 
approximately\cite{hsu,rand} as we reproduce here. From

\begin{equation}
b_{{\vec k} \uparrow}^\dagger | \Phi \rangle
= \frac{1}{u_{\vec k}} c_{ {\vec k} \uparrow}^\dagger | \Phi \rangle
= \frac{1}{v_{\vec k}} c_{- {\vec k} \downarrow} | \Phi \rangle
\end{equation}

\noindent we get

\begin{displaymath}
\frac{1}{N} \sum_{\vec k}^N u_k^2
\frac{\langle \Phi | b_{{\vec k} \uparrow} \Pi_\alpha^2
b_{{\vec k} \uparrow}^\dagger | \Phi \rangle }
{ \langle \Phi | \Pi_\alpha^2 | \Phi \rangle }
= \frac{\langle \Phi | c_{j \uparrow} \Pi_\alpha^2
c_{j \uparrow}^\dagger | \Phi \rangle }
{\langle \Phi | \Pi_\alpha^2 | \Phi \rangle }
\end{displaymath}

\begin{equation}
= z_0 \biggl[ \frac{1 + (1 - \alpha) z_0 z}{1 + 2 z z_0 + (1 - \alpha )
( z z_0)^2}\biggr] = \frac{1 + \delta}{2}
\end{equation}

\noindent
and

\begin{displaymath}
\frac{1}{N} \sum_{\vec k}^N v_k^2
\frac{\langle \Phi | b_{{\vec k} \uparrow} \Pi_\alpha^2
b_{{\vec k} \uparrow}^\dagger | \Phi \rangle }
{ \langle \Phi | \Pi_\alpha^2 | \Phi \rangle }
= \frac{\langle \Phi | c_{j \uparrow}^\dagger \Pi_\alpha^2
c_{j \uparrow} | \Phi \rangle }{\langle \Phi | \Pi_\alpha^2 | \Phi \rangle }
\end{displaymath}

\begin{equation}
= \frac{1}{z_0} \biggl[ \frac{z z_0 + (z z_0)^2}{1 + 2 z z_0 + (1 - \alpha ) 
(z z_0)^2} \biggr] = \frac{1 - \delta}{2}
\; \; \; .
\end{equation}

\noindent Since these are the same values as the unprojected quantities 
(\ref{coh}), we see that 
\beq
\frac{ \langle \Psi | \Psi \rangle}
{\langle {\vec k} \sigma | {\vec k} \sigma  \rangle} \simeq 1
\eneq

We  now consider the low-energy spectroscopic properties of the Gossamer 
superconductor. We will repeatedly use tge matrix elements for $j \neq j'$. In
order to evaluate them, we assume that the amplitude for a given
configuration is weighted by the square root of its corresponding
probability, and that these weights add equally. We get

\begin{displaymath}
\frac{\langle \Phi | c_{j \uparrow} \Pi_\alpha^2
c_{j' \uparrow}^\dagger | \Phi \rangle }
{\langle \Phi | \Pi_\alpha^2 | \Phi \rangle } \times
\frac{ \langle \Phi | \Phi \rangle }
{\langle \Phi | c_{j \uparrow} c_{j' \uparrow}^\dagger | \Phi \rangle }
\end{displaymath}

\begin{equation}
\simeq \frac{4}{1 - \delta^2} (z z_0) \biggl[
\frac{1 + (1 - \alpha) z z_0}{1 + 2 z z_0 + (1 - \alpha ) (z z_0)^2} \biggr]^2 
\end{equation}

\noindent
and

\begin{displaymath}
\frac{\langle \Phi | c_{j \uparrow}^\dagger \Pi_\alpha^2
c_{j' \uparrow} | \Phi \rangle }
{\langle \Phi | \Pi_\alpha^2 | \Phi \rangle } \times
\frac{ \langle \Phi | \Phi \rangle }
{\langle \Phi | c_{j \uparrow}^\dagger c_{j' \uparrow} | \Phi \rangle }
\end{displaymath}

\begin{equation}
\simeq \frac{4}{1 - \delta^2} (\frac{z}{z_0}) \biggl[
\frac{1 + z z_0}{1 + 2 z z_0 + (1 - \alpha ) (z z_0)^2} \biggr]^2 
\; \; \; .
\end{equation}

We thus find the matrix element for photoemission and inverse photoemission
to be

\begin{displaymath}
\frac{ \langle - {\vec k} \downarrow | c_{{\vec k} \uparrow} |
\Psi \rangle }
{\sqrt{\langle - {\vec k} \downarrow | - {\vec k} \downarrow \rangle \;
\langle \Psi | \Psi \rangle}} = g v_{\vec k}
\; \; \; ,
\end{displaymath}

\begin{equation}
\frac{ \langle - {\vec k} \downarrow | c_{{\vec k} \uparrow}^\dagger |
\Psi \rangle }
{\sqrt{\langle - {\vec k} \downarrow | - {\vec k} \downarrow \rangle \;
\langle \Psi | \Psi \rangle}} = g u_{\vec k}
\; \; \; ,
\end{equation}

\noindent
where

\begin{equation}
g^2 \simeq \frac{2 \alpha_0}{\alpha} \biggl\{ 1 - \frac{\alpha_0}{\alpha}
\biggl[ \frac{1 - \sqrt{1 - \alpha ( 1 - \delta^2)}}
{1 - \delta^2} \biggr] \biggr\}
\; \; \; .
\end{equation}

\noindent Photoemission intensity is suppresed. We can estimate the superfluid 
density and see a similar suppression:

\begin{equation}
\frac{< \! \Psi | c_{j \uparrow } c_{j' \downarrow} | \Psi \! >}
{< \! \Psi | \Psi \! >} \simeq g^2
\frac{< \! \Phi | c_{j \uparrow } c_{j' \downarrow} | \Phi \! >}
{< \! \Phi | \Phi \! >}
\; \; \; .
\end{equation}

\noindent Thus under strong projection near half-filling this model exhibits 
the pseudogap phenomenon: The quasiparticle dispersion remains unchanged from
its unperturbed value $E_{\vec k}$ as $\alpha_0$ increases from 0 to 1, but
the superfluid density decreases from 1 to $2|\delta|/(1 + |\delta|)$.
The strong projector collapses the superfluid density with doping
and introduces correlations intrinsic to an antiferromagnetic insulator as
Hubbard band-liek lobes grow as we shall show next.

\section{Magnetic Correlations of the Gossamer Superconductor} 

In the present and the following section we consider how the Gossamer 
superconductor under strong projection and near half-filling develops behavior 
that is practically impossible to differentiate from that of a correlated 
insulator. We now study the Gossamer magnetic behavior.

In order to determine the electronic correlations arising in the
Gossamer Hamiltonian Eq. (\ref{gossham}),
we expand it and analyze its terms:

\beq {\cal{H}} =
\sum_{\vec{k} \sigma} E_{\vec{k}} B^\dagger_{\vec{k} \sigma}
B_{\vec{k} \sigma} = \cal{A} + \cal{B} + \cal{C}
\eneq \noindent
where $\cal{A}, \cal{B}, \cal{C}$ are, explicitly:
\begin{displaymath} {\cal{A}}= \sum_{\vec{k}} \frac{E_{\vec{k}}}{N} 
\sum_{i,j}^N e^{-i \vec{k}
(\vec{r}_i -\vec{r}_j)} \{z_0^{-1} u_{\vec{k}}^2 (1+\beta_0
n_{i\downarrow})(1+\beta_0 n_{j\downarrow}) c^\dagger_{i \uparrow}
c_{j \uparrow} +
\end{displaymath}
\begin{equation}
z_0 v_{\vec{k}}^2 (1-\alpha_0 n_{i \downarrow})(1-\alpha_0 n_{j
\downarrow}) c_{i \uparrow} c^\dagger_{j \uparrow}\} +
\{\uparrow\rightleftarrows\downarrow \}
\end{equation}

\begin{displaymath}
{\cal{B}}=\sum_{\vec{k}} \frac{E_{\vec{k}}}{N} \sum_{i,j}^N
e^{-i\vec{k}(\vec{r}_i - \vec{r}_j)} u_{\vec{k}} v_{\vec{k}} \{
(1+\beta_0 n_{i \downarrow})(1-\alpha_0 n_{j \uparrow})
c^\dagger_{i \uparrow} c^\dagger_{j\downarrow} -
\end{displaymath}
\begin{equation}
+(1+\beta_0 n_{j \downarrow})(1-\alpha_0 n_{i \uparrow}) c_{i
\downarrow} c_{j\uparrow} \}-\{\uparrow\rightleftarrows\downarrow
\} \label{B}
\end{equation}

\begin{displaymath}
{\cal{C}}=\sum_{\vec{k}} \frac{E_{\vec{k}}}{N}  \sum_{j}^N
u_{\vec{k}} v_{\vec{k}} \{ \alpha_0 (1+\beta_0 n_{j \downarrow})
c^\dagger_{j \uparrow} c^\dagger_{j \downarrow}
\end{displaymath}

\begin{equation}
+ \beta_0 (1-\alpha_0 n_{j \uparrow}) c_{j \downarrow} c_{j
\uparrow} \} - \{\uparrow\rightleftarrows\downarrow \}
\end{equation}

\noindent The term $\cal{C}$ vanishes the identity 

\beq
E_{\vec{k}}
u_{\vec{k}} v_{\vec{k}} = \frac{\Delta_{\vec{k}}}{2}
\eneq

\noindent and the relation:

\beq \sum_{\vec{k}} e^{-i\vec{k}(\vec{r}_i - \vec{r}_j)}
\Delta_{\vec{k}} = \left\{
\begin{array}{ccc}
  0 & , & \vec{r}_i \ne \vec{r}_j  +\vec{a}\\
  \Delta_0 & , & \vec{a} \parallel \hat{x} \\
  -\Delta_0 & , & \vec{a} \parallel \hat{y} \\
\end{array} \right\}\eneq 

\noindent (true for a d-wave gap) where $\vec{a}$ is vector pointing toward
a nearest neighbor in the lattice. This term would be nonzero for an s-wave 
superconductor and it might be interesting to study what happens in such a 
case. The $\cal{A}$ term is responsible for the chemical potential, kinetic 
energy, as well as a Hubbard U terms. $\cal{B}$ is responsible for the
super-conducting part of the Gossamer Hamiltonian. The
d-wave form of the gap makes $\cal{B}$ have no on-site
contributions, but it may be of interest to look at an s-wave superconductor. 
$\cal{A}$ contains on-site and off-site contributions:

\begin{displaymath}
{\cal{A}} = {\cal{A}}_{on
\; site} + {\cal{A}}_{off\;site}
\end{displaymath}

\begin{displaymath} {\cal{A}}_{on
\; site}= \sum_{\vec{k}} \frac{E_{\vec{k}}}{N} \sum_{j}^N  \{z_0^{-1}
u_{\vec{k}}^2
(1+\beta_0 n_{j\downarrow})^2 c^\dagger_{j \uparrow} c_{j \uparrow} +
\end{displaymath}

\begin{equation} + z_0 v_{\vec{k}}^2 (1-\alpha_0 n_{j \downarrow})^2
c_{j \uparrow} c^\dagger_{j \uparrow} \} +\{ \uparrow
\rightleftarrows \downarrow \}
\end{equation} \noindent

\begin{displaymath} {\cal{A}}_{off\;site}
=\sum_{\vec{k}} \frac{E_{\vec{k}}}{N} \sum_{i \ne j}^N e^{-i
\vec{k} (\vec{r}_i -\vec{r}_j)}
\end{displaymath}

\begin{displaymath}
\times \{z_0^{-1} u_{\vec{k}}^2 (1+\beta_0
n_{i\downarrow})(1+\beta_0 n_{j\downarrow}) c^\dagger_{i \uparrow}
c_{j \uparrow} +
\end{displaymath}

\begin{equation}
+z_0 v_{\vec{k}}^2 (1-\alpha_0 n_{i
\downarrow})(1-\alpha_0 n_{j \downarrow})  c^\dagger_{j \uparrow}
c_{i \uparrow} \} + \{\uparrow \rightleftarrows \downarrow \}
\end{equation}

\noindent The Hubbard U term arises from 
${\cal{A}}_{on \; site}$. which can be written as 

\begin{displaymath} {\cal{A}}_{on
\; site}
= \sum_{\vec{k}} \frac{E_{\vec{k}}}{N} \sum_j^N \{ 2 z_0
v^2_{\vec{k}} +
\end{displaymath}

\begin{displaymath}
+[z_0^{-1} u^2_{\vec{k}} - z_0 v^2_{\vec{k}} -2\alpha_0 z_0
v^2_{\vec{k}} + \alpha_0^2 z_0 v^2_{\vec{k}}] (n_{j \uparrow} +
n_{j \downarrow}) +
\end{displaymath}

\beq +[z_0^{-1} u^2_{\vec{k}}(4\beta_0 + 2 \beta_0^2) + z_0
v^2_{\vec{k}}(4\alpha_0 - 2\alpha_0^2)]n_{j \uparrow}  n_{j
\downarrow}\}
\eneq

\noindent The first term is a flucutation energy, the second term
is a chemical potential, and the third term
is the Hubbard U term ($\sum_j^N U n_{j \uparrow}  n_{j \downarrow} $)
with:

\begin{equation}
U=\sum_{\vec{k}} \frac{E_{\vec{k}}}{N} [z_0^{-1}
u^2_{\vec{k}}(4\beta_0 + 2 \beta_0^2) + z_0
v^2_{\vec{k}}(4\alpha_0 - 2\alpha_0^2)]
\end{equation}

\noindent Thus the Gossamer Hamiltonian has a Hubbard $U$ term which arises 
from the ``projector''..

The Gossamer Hamiltonian is constructed such that its ground state is 
superconducting for all nonzero dopings, but It will become increasingly 
susceptible to antiferromagnetism at zero doping under almost full projection
where the superfluid density is arbitrarily collapses to zero. Concentrating 
on half-filling $\delta =0$, $U$ is given by: 

\beq 
U|_{\delta = 0}= \sum_{\vec{k}} \frac{2
E_{\vec{k}}}{N} \frac{\alpha_0(2- \alpha_0)}{1-\alpha_0} \label{U}
\eneq 

\noindent At almost full projection $\alpha_0 \rightarrow 1^-$, $U$ grow 
arbitrarily large.

The off-site contributions of $\cal{A}$ provide the hopping (kinetic energy)
term in the Hamiltonian. After partial projection, particularizing to zero 
doping and {\it imposing} the mean field values 
$\langle n_{i\uparrow} \rangle = \langle n_{i\downarrow} \rangle = 1/2 $, the 
off-site contribution becomes:

\beq {\cal{A}}_{off\;site} =\frac{1}{4}
\frac{(2-\alpha_0)^2}{(1-\alpha_0)} \sum_{\vec{k} \sigma}
(\epsilon_{\vec{k}} - \mu) c^\dagger_{\vec{k} \sigma} c_{\vec{k}
\sigma}  \label{t}
\eneq

\noindent At half filling, the kinetic energy term of the Gossamer Hamiltonian 
is renormalized by a constant factor that grows arbitrarily large as we
go to full projection. On the other hand, upon strong projection, the 
physically relevant ratio, $U/t$, remains finite and approaches a 
number of order unity or greater which provides the right physics for the 
appearance of strong antiferromagnetic correlations and insulation\cite{zg2}. 

The superconducting part of the Gossamer Hamiltonian, $\cal{B}$ when 
unprojected, just the Cooper pairing  attraction term
\beq
\sum_{\vec{k}} \Delta_{\vec{k}} [
c^{\dagger}_{\vec{k} \uparrow} c^{\dagger}_{-\vec{k} \downarrow} +
c_{-\vec{k} \downarrow} c_{\vec{k} \uparrow} ]
\eneq 

\noindent At half-filling and {\it imposing} the mean field condition
$\langle n_{i\uparrow} \rangle = \langle n_{i \downarrow} \rangle =1/2$, and
keeping in mind that $E_{\vec{k}}$ and $\Delta_{\vec{k}}$ are even in $\vec{k}$
we thus obtain

\beq {\cal{B}}=\frac{1}{4}
\frac{(2-\alpha_0)^2}{1-\alpha_0} \sum_{\vec{k}} \Delta_{\vec{k}}
 (c^\dagger_{\vec{k} \uparrow}
c^\dagger_{-\vec{k} \downarrow} + c_{-\vec{k} \downarrow}
c_{\vec{k} \uparrow})
\eneq

\noindent The superconducting gap survives upon projection, but is renormalized
by the same constant as the kinetic energy. 
Upon strong projection, the physically relevant ratio, $U/ \Delta_0$, is a 
number of order unity or greater, the right physics for antiferromagnetism and 
insulation. It is  very interesting that the gap survives 
along with Hubbard $U$ term at half-filling where the superfluid density can
be arbitrarily close to zero upon strong projection.

At half-filling and under strong projection the Gossamer superconductor 
Hamiltonian is thus a Hubbard Hamiltonian with a d-wave pairing interaction 
added to it. We now define the ``spinors''

\begin{equation}
\Psi_{\vec{k}} \equiv \left[ \begin{array}{c}
c_{\vec{k} \uparrow} \\ c_{-\vec{k} \downarrow}^\dagger \end{array} \right]
\; ,
\end{equation}

\noindent so that the ``noninteracting'' part of the Gossamer Hamiltonian, 
i.e. the part with the $U$ term disregarded, can be written

\begin{equation}
\text{$\cal{H}$} = \frac{1}{4}\frac{(2-\alpha_0)^2}{1-\alpha_0} \left[
\begin{array}{cc}
\epsilon_{\bf k}  & \Delta_{\bf k} \\
\Delta_{\bf k} &  - \epsilon_{\bf k}
\end{array} \right]
\end{equation}

\noindent where $\mu$ has been omitted because we are at half-filling.
The bare Green function, $G_{\bf k}(E)=1/(E - \text{$\cal{H}$})$, is then 
given by

\begin{equation}
G_{\bf k}(E)= \frac{1}{E^2 - \gamma^2 (\epsilon_{\bf k}^2
+ \Delta_{\bf k}^2 )} \left[ \begin{array}{cc}
E + \gamma \epsilon_{\bf k}  & - \gamma \Delta_{\bf k} \\
- \gamma \Delta_{\bf k} &  E - \gamma \epsilon_{\bf k}
\end{array} \right]
\end{equation}

\noindent with $\gamma \equiv (2-\alpha_0)^2 / 4 (1-\alpha_0)$.

\begin{figure}
\input{test.ltx}
\caption{Dependence of the imaginary part of the spin susceptibility in
RPA approximation on the energy in units of $t$. The specific curves
plotted here are for $\Delta_0 = 0.4 t$ and $q=(\pi,\pi)$. Upon
increasing the Hubbard U toward the critical value U$=1.43t$ we notice the
divergence of the susceptibility, a sign that magnetic order is about to
set in.} \label{fig}
\end{figure}
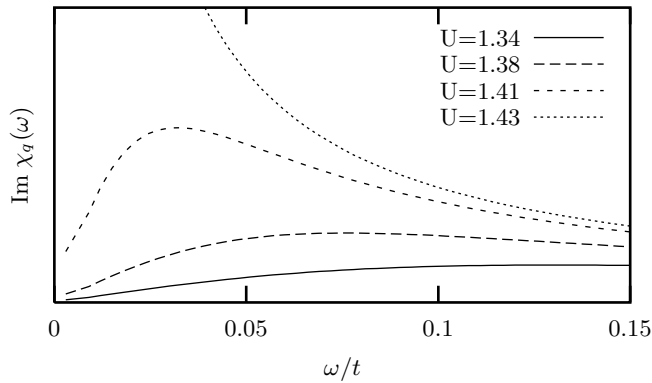

In order to show the magnetic ordering properties of the Gossamer 
Hamiltonian
at half-filling, we will compute the magnetic susceptibility and tune it
through the transition. The bare susceptibility is given by

\begin{equation}
\chi_q^0 (\omega) = \frac{1}{(2\pi)^3} \int \int
{\rm Tr} [ G_k (E) G_{k+q}(E + \omega) ] \; dE dk \; \; .
\end{equation}

We calculate the effects of $U$ by the the ladder approximation for 
the the spin susceptibility $\chi_q (\omega) = \chi_q^0 (\omega)/ [1 + U 
\chi_q^0 (\omega )]$. The numerical evaluation of the spin susceptibility is 
shown in the figure\cite{us}. We see that beyond at critical value for $U$ of 
order of $t$ and or$\Delta_0$, the system goes to the critical point becoming 
infinitely susceptible to going over into an antiferromagnetic insulator as 
signaled by the diverging susceptibility at the critical value. The ladder 
technique cannot provide the correct critical value of $U/t$, nor the correct 
critical exponents, but it will, however, provide a faithful qualitative 
picture of the transition, and of the divergence of the spin susceptibility at 
the critical point for the development of antiferromagnetic order. 

\section{Mott-Hubbard Insulating-like Behavior of the Gossamer Superconductor}

Consistent with the approach to antiferromagnetic order at half-filling, we
will see the formation of the Mott-Hubbard gap. Quasiparticle photoemission
can only account for a small part of the sum rule

\begin{equation}
\frac{\langle \Psi | c_{{\vec k} \sigma}^\dagger c_{{\vec k} \sigma}
| \Psi \rangle }{\langle \Psi | \Psi \rangle } \simeq g^2 v_{\vec k}^2 +
 \frac{1- g^2}{2}
\; \; \; .
\end{equation}

\noindent
The rest must occur at a higher energy scale, the value of which may
be estimated by computing the expected energy of a hole. Using the 
anticommutators 

\begin{equation}
\{ B_{{\vec k} \uparrow} , c_{j \uparrow} \}
= \frac{1}{\sqrt{N}} e^{i {\vec k} \cdot {\vec r}_j} \biggl[
- z_0^{1/2} \alpha_0 v_{\vec k}
c_{j \uparrow} c_{j \downarrow}^\dagger \biggr]
\end{equation}

\begin{displaymath}
\{ B_{{\vec k} \downarrow} , c_{j \uparrow} \}
= \frac{1}{\sqrt{N}} e^{i {\vec k} \cdot {\vec r}_j} \biggl[
- z_0^{-1/2} \beta_0 u_{\vec k}
c_{j \uparrow} c_{j \downarrow}
\end{displaymath}

\begin{equation}
+ z_0^{1/2} v_{\vec k} ( 1 - \alpha_0 n_{j \downarrow} ) \biggr]
\end{equation}

\noindent it follows that

\begin{displaymath}
\langle \Psi | c_{{\vec k} \sigma}^\dagger {\cal H} c_{{\vec k} \sigma} |
\Psi \rangle = z_0 \biggl[ 1 - \alpha_0 \frac{1 - \delta}{2} \biggr]^2
v_{\vec k}^2 E_{\vec k}
\end{displaymath}

\begin{equation}
+ \frac{1}{N} \sum_{\vec q}^N E_{{\vec k} + {\vec q}} \biggl[
z_0 \alpha_0^2 A_{\vec q} v_{{\vec k} + {\vec q}}^2 +
\frac{\beta_0^2}{\alpha_0} B_{\vec q} u_{{\vec k} + {\vec q}}^2 \biggr]
\; \; \; ,
\end{equation}

\noindent with 

\begin{displaymath}
A_{\vec q} = \sum_j^N \biggl[
\frac{\langle \Psi | {\vec S}_0 \cdot {\vec S}_j | \Psi \rangle }
{\langle \Psi | \Psi \rangle }
\end{displaymath}

\begin{equation}
+ \frac{1}{4} \frac{\langle \Psi | n_0 n_j | \Psi \rangle }
{\langle \Psi | \Psi \rangle} - \frac{(1 - \delta)^2}{4}
\biggr] e^{i {\vec q} \cdot {\vec r}_j}
\end{equation}

\begin{equation}
B_{\vec q} = \sum_j^N
\frac{\langle \Psi | c_{0 \uparrow}^\dagger c_{0 \downarrow}^\dagger
c_{j \downarrow} c_{j \uparrow} | \Psi \rangle }
{\langle \Psi | \Psi \rangle } e^{i {\vec q} \cdot {\vec r}_j}
\; \; \; .
\end{equation}

\noindent If we go through similar gymnastics we determine that the energy to 
add an electron is the exact
particle-hole conjugate of this expression, produced from it by
interchanging $u_{\bf k}$ and $v_{\bf k}$, negating $\delta$, and
substituting of $1 - n_j$ for $n_j$.  Therefore the electron spectral function 
is symmetric at half-filling.  We see that, in the Gossamer superconductor, the
spectral function consists of Mott-Hubbard ``lobes'' at high energies with a 
faint band of states at mid-gap associated with the Gossamer quasiparticles. 
The chemical potential stays pinned at midgap with doping. 

We have seen that, under strong projection ($\alpha_0 \rightarrow 1$), the 
Gossamer superconductor
has a superfluid density that collapses with doping and projection. This 
collapsing superfluid density leads to a temperature order parameter phase 
instability\cite{emery} consistent with the transition out of the 
superconducting state in underdoped cuprates\cite{timusk}. Even without the 
development of antiferromagnetism, such a superconductor would be insulating 
since it would dephase due to the small superfluid density\cite{dynes} perhaps
thorugh lattice-mediated crystallization of the gossamer superconductor which
would also provide a ready explanation for why the
static stripes\cite{tranquada} occur at $\delta = 1/8$ in some
cuprates and not others, why stripes are destroyed by moderate pressure
\cite{pressure}, and why the materials insulate when subjected to
strong magnetic fields\cite{boebinger}. 

\section{metal}

The Gossamer technique should be applied with more care to a metallic Fermi sea
ground state than to a superconducting BCS ground state. The metal has 
the abundance of low energy degrees of freedom which might make the
projection uncontrolled in the infrared leading to unphysical
results. On the other hand, the superconductor does not have such a plethora of
low energy degrees of freedom and the calculation is assured to have no 
infrared problems: it is regularized by the superconducting order. 
After obtaining the results, we collapse the gap to zero to study the physics 
of Gossamer metals.

Since the Gossamer superconducting ground state is adiabatically continuable to
the BCS ground state by continuously varying $\alpha_0$ to zero, in a similar 
fashion once we collapse the gap to obtain the Gossamer metal ground state, it 
will be adiabatically deformable to a regular Fermi sea metallic ground state
and, hence they will be the same zero temperature phase of matter.

In order to study quasiparticle dispersion in the Gossamer metal, we remember
that, for the superconductor, the wavefunction 
\beq
| {\vec k} \sigma \rangle = \Pi_\alpha b_{{\vec k}
\sigma}^\dagger | \Phi \rangle
\eneq

\noindent represents an appropriate approximation to the low energy 
quasiparticle excitations with dispersion

\begin{equation}
\frac{\langle {\vec k} \sigma | {\cal H} | {\vec k} \sigma \rangle }
{ \langle {\vec k} \sigma | {\vec k} \sigma \rangle }
= E_{\vec k} \; \frac{ \langle \Psi | \Psi \rangle}
{\langle {\vec k} \sigma | {\vec k} \sigma  \rangle }
\simeq E_k 
\end{equation}

\noindent that, after collapsing the superconducting gap, is unchanged from the
dispersion from that in  the regular metal. This is contrary to usual 
thinking in which correlation effects are believed to make the charge carriers
heavy\cite{heavy}. In the Gossamer metal, the carriers are {\it not}
getting arbitrarily heavy as we approach the transition, but the
metallic band is becoming thinner and the missing spectral weight goes
to energies far from the Fermi sea to forming Hubbard bands. The
carriers are just as fast, there are just less of them which degrades
the conductivity.

The photoemission amplitudes were calculated for the Gossamer
superconductor. Collapsing the gap, we obatin them for the Gossamer metal:

\begin{displaymath}
\frac{ < \! - {\vec k} \downarrow | c_{{\vec k} \uparrow} |
\Psi \! > }
{\sqrt{< \! - {\vec k} \downarrow | - {\vec k} \downarrow > \;
< \! \Psi | \Psi \! > }} = 0
\; \; \; ,
\end{displaymath}
\begin{equation}
\frac{ < \! - {\vec k} \downarrow | c_{{\vec k} \uparrow}^\dagger |
\Psi \! > }
{\sqrt{< \! - {\vec k} \downarrow | - {\vec k} \downarrow > \;
< \! \Psi | \Psi \! > }} = g
\; \; \; ,
\end{equation}

\noindent with 

\begin{equation}
g^2 \simeq \frac{2 \alpha_0}{\alpha} \biggl\{ 1 - \frac{\alpha_0}{\alpha}
\biggl[ \frac{1 - \sqrt{1 - \alpha ( 1 - \delta^2)}}
{1 - \delta^2} \biggr] \biggr\}
\; \; \; .
\end{equation}

\noindent The suppression of photoemission amplitude is caused by the smaller 
number of metallic electrons whose number is diminished from that in the 
unprojected metal by a factor $g^2$ which goes to 
$2|\delta|/(1 + |\delta|)$ as $\alpha_0 
\rightarrow 1$. This is consistent with the superfluid density suppression in 
the Gossamer superconductor which goes like $g^2$ as there is a
sum rule for the superconductor making the density of conducting electrons 
equal to the superfluid density. 

The existence of the growing Hubbard U term means that as we go to
half-filling and full projection magnetic correlations will get
enhanced leading to a diverging magnetic susceptibility in the exact
same way as for the Gossamer superconductor  after we collapse
the gap. The spectral weight will consist of Mott-Hubbard bands at
high energies with the chemical potential pinned at midgap at an ever
fainter band from which the Gossamer quasiparticles are excited with
a dispersion unchanged from the noninteracting metal. 

\section{Discussion}

We have reviewed the ideas of the Gossmaer technique for superconductors an
the recent extensions to metals. The technique has the advantage of introducing
strong correlations, but yet prohibits the ground state for undegoing a 
transition to the insualting state. The best we can do is get to the critical 
point at half-filling by projecting fully. 

For the superconductor we saw that the superfluid density and photoemssion 
amplitude becomes suppressed at half-filling and the magnetic correlations 
become enhanced as we tune the correlations up. Thus the superconducting state 
with strong on-site repulsion is unstable toward insulation and 
antiferromagnetism close to half-filling, being arbitrarily close to a 
continuous zero temperature phase transition into an antiferromagnetic 
insulator. The critical point is at half-filling when fully projected and it is
the Anderson RVB ground state\cite{rvb}. For the Gossamer superconductor the 
instability is exactly at half-filling while for a different Hamiltonian the 
instability can occur at nonzero doping. Since the Gossamer superconductor is 
adiabatically continuable to a completely regular BCS superconductor our 
correlation effects are generic to the superconducting state. 

The proximity to the antiferromagnetic transition found here under strong
projection will make the spectroscopic properties of the material be very much 
like those of an antiferromagnetic insulator near half-filling. The superfluid
density will be so low that it would be almost impossible to tell that the 
system is not an insulator except at extremely long wavelengths or low energy 
scales. An antiferromagnet with a small interpenetrating density of dephased 
superfluid
provides a possible explanation for the recent measurements of metallic 
transport below the N\'{e}el temperature in underdoped LSCO\cite{mottcond}. 
That the charge mobility in these measurements is equal to that in the 
optimally doped material\cite{ando} suggests a common origin, possibly the 
dephased Gossamer superconductor. Moreover, adding by hand an extra Hubbard 
term, an insulating static stripe phase would be stabilized with a possible 
coexistence of dephased superfluid. Coupling of the coexisting dephased 
superfluid to the stripe phase would lead to anisotropic 
Copper-Oxygen plane charge transport\cite{ando0}. 

Phase fluctuations have not been included in the Gossamer superconducting 
Hamiltonian because they are irrelevant to the fermi spectrum, which is
characterized by an energy scale much higher than the superconducting
T$_c$.  However, they are essential for accounting for both the Uemura
plot and strange-metal transport above T$_c$. The transport of a dephased 
fermion fluid with Cooper pair correaltions formed when the order parameter 
dephases would not exhibit any traditional metallic behavior.  

We also reviewes the applciation of the Gossamer technique to metals. The 
Gossamer metal describes strongly correlated bad metal
behavior that is very hard to distinguish from true insulating
behavior, the implication being that the magnetically disordered
insulating-like phases in some systems might really be a Gossamer metal
with very much degraded conductivity. The degraded conductivity
arrises from a depletion of spectral weight of metallic electrons and
not by an ever growing effective mass of the carriers. 

{\bf Acknowledgements} B.A.B. was supported by the Stanford 
Graduate Fellowship and DOE contract No. DE-AC03-76SF00515. D.I.S. was 
supported by NASA grant NAS 8-39225 to Gravity Probe B.

\end{document}

%% file: test.ltx
% GNUPLOT: plain TeX with Postscript
\begingroup
  \catcode`\@=11\relax
  \def\GNUPLOTspecial{%
    \def\do##1{\catcode`##1=12\relax}\dospecials
    \catcode`\{=1\catcode`\}=2\catcode\%=14\relax\special}%
\expandafter\ifx\csname GNUPLOTpicture\endcsname\relax
  \csname newdimen\endcsname\GNUPLOTunit
  \gdef\GNUPLOTpicture(#1,#2){\vbox to#2\GNUPLOTunit\bgroup
    \def\put(##1,##2)##3{\unskip\raise##2\GNUPLOTunit
      \hbox to0pt{\kern##1\GNUPLOTunit ##3\hss}\ignorespaces}%
    \def\ljust##1{\vbox to0pt{\vss\hbox to0pt{##1\hss}\vss}}%
    \def\cjust##1{\vbox to0pt{\vss\hbox to0pt{\hss ##1\hss}\vss}}%
    \def\rjust##1{\vbox to0pt{\vss\hbox to0pt{\hss ##1}\vss}}%
    \def\stack##1{\let\\=\cr\tabskip=0pt\halign{\hfil ####\hfil\cr ##1\crcr}}%
    \def\lstack##1{\hbox to0pt{\vbox to0pt{\vss\stack{##1}}\hss}}%
    \def\cstack##1{\hbox to0pt{\hss\vbox to0pt{\vss\stack{##1}}\hss}}%
    \def\rstack##1{\hbox to0pt{\vbox to0pt{\stack{##1}\vss}\hss}}%
    \vss\hbox to#1\GNUPLOTunit\bgroup\ignorespaces}%
  \gdef\endGNUPLOTpicture{\hss\egroup\egroup}%
\fi
\GNUPLOTunit=0.1bp
{\GNUPLOTspecial{!
%!PS-Adobe-2.0 EPSF-2.0
%%Title: test.ltx
%%Creator: gnuplot 3.7 patchlevel 1
%%CreationDate: Mon Feb  3 18:46:45 2003
%%DocumentFonts: 
%%BoundingBox: 0 0 252 151
%%Orientation: Landscape
%%EndComments
/gnudict 256 dict def
gnudict begin
/Color false def
/Solid false def
/gnulinewidth 5.000 def
/userlinewidth gnulinewidth def
/vshift -33 def
/dl {10 mul} def
/hpt_ 31.5 def
/vpt_ 31.5 def
/hpt hpt_ def
/vpt vpt_ def
/M {moveto} bind def
/L {lineto} bind def
/R {rmoveto} bind def
/V {rlineto} bind def
/vpt2 vpt 2 mul def
/hpt2 hpt 2 mul def
/Lshow { currentpoint stroke M
  0 vshift R show } def
/Rshow { currentpoint stroke M
  dup stringwidth pop neg vshift R show } def
/Cshow { currentpoint stroke M
  dup stringwidth pop -2 div vshift R show } def
/UP { dup vpt_ mul /vpt exch def hpt_ mul /hpt exch def
  /hpt2 hpt 2 mul def /vpt2 vpt 2 mul def } def
/DL { Color {setrgbcolor Solid {pop []} if 0 setdash }
 {pop pop pop Solid {pop []} if 0 setdash} ifelse } def
/BL { stroke userlinewidth 2 mul setlinewidth } def
/AL { stroke userlinewidth 2 div setlinewidth } def
/UL { dup gnulinewidth mul /userlinewidth exch def
      10 mul /udl exch def } def
/PL { stroke userlinewidth setlinewidth } def
/LTb { BL [] 0 0 0 DL } def
/LTa { AL [1 udl mul 2 udl mul] 0 setdash 0 0 0 setrgbcolor } def
/LT0 { PL [] 1 0 0 DL } def
/LT1 { PL [4 dl 2 dl] 0 1 0 DL } def
/LT2 { PL [2 dl 3 dl] 0 0 1 DL } def
/LT3 { PL [1 dl 1.5 dl] 1 0 1 DL } def
/LT4 { PL [5 dl 2 dl 1 dl 2 dl] 0 1 1 DL } def
/LT5 { PL [4 dl 3 dl 1 dl 3 dl] 1 1 0 DL } def
/LT6 { PL [2 dl 2 dl 2 dl 4 dl] 0 0 0 DL } def
/LT7 { PL [2 dl 2 dl 2 dl 2 dl 2 dl 4 dl] 1 0.3 0 DL } def
/LT8 { PL [2 dl 2 dl 2 dl 2 dl 2 dl 2 dl 2 dl 4 dl] 0.5 0.5 0.5 DL } def
/Pnt { stroke [] 0 setdash
   gsave 1 setlinecap M 0 0 V stroke grestore } def
/Dia { stroke [] 0 setdash 2 copy vpt add M
  hpt neg vpt neg V hpt vpt neg V
  hpt vpt V hpt neg vpt V closepath stroke
  Pnt } def
/Pls { stroke [] 0 setdash vpt sub M 0 vpt2 V
  currentpoint stroke M
  hpt neg vpt neg R hpt2 0 V stroke
  } def
/Box { stroke [] 0 setdash 2 copy exch hpt sub exch vpt add M
  0 vpt2 neg V hpt2 0 V 0 vpt2 V
  hpt2 neg 0 V closepath stroke
  Pnt } def
/Crs { stroke [] 0 setdash exch hpt sub exch vpt add M
  hpt2 vpt2 neg V currentpoint stroke M
  hpt2 neg 0 R hpt2 vpt2 V stroke } def
/TriU { stroke [] 0 setdash 2 copy vpt 1.12 mul add M
  hpt neg vpt -1.62 mul V
  hpt 2 mul 0 V
  hpt neg vpt 1.62 mul V closepath stroke
  Pnt  } def
/Star { 2 copy Pls Crs } def
/BoxF { stroke [] 0 setdash exch hpt sub exch vpt add M
  0 vpt2 neg V  hpt2 0 V  0 vpt2 V
  hpt2 neg 0 V  closepath fill } def
/TriUF { stroke [] 0 setdash vpt 1.12 mul add M
  hpt neg vpt -1.62 mul V
  hpt 2 mul 0 V
  hpt neg vpt 1.62 mul V closepath fill } def
/TriD { stroke [] 0 setdash 2 copy vpt 1.12 mul sub M
  hpt neg vpt 1.62 mul V
  hpt 2 mul 0 V
  hpt neg vpt -1.62 mul V closepath stroke
  Pnt  } def
/TriDF { stroke [] 0 setdash vpt 1.12 mul sub M
  hpt neg vpt 1.62 mul V
  hpt 2 mul 0 V
  hpt neg vpt -1.62 mul V closepath fill} def
/DiaF { stroke [] 0 setdash vpt add M
  hpt neg vpt neg V hpt vpt neg V
  hpt vpt V hpt neg vpt V closepath fill } def
/Pent { stroke [] 0 setdash 2 copy gsave
  translate 0 hpt M 4 {72 rotate 0 hpt L} repeat
  closepath stroke grestore Pnt } def
/PentF { stroke [] 0 setdash gsave
  translate 0 hpt M 4 {72 rotate 0 hpt L} repeat
  closepath fill grestore } def
/Circle { stroke [] 0 setdash 2 copy
  hpt 0 360 arc stroke Pnt } def
/CircleF { stroke [] 0 setdash hpt 0 360 arc fill } def
/C0 { BL [] 0 setdash 2 copy moveto vpt 90 450  arc } bind def
/C1 { BL [] 0 setdash 2 copy        moveto
       2 copy  vpt 0 90 arc closepath fill
               vpt 0 360 arc closepath } bind def
/C2 { BL [] 0 setdash 2 copy moveto
       2 copy  vpt 90 180 arc closepath fill
               vpt 0 360 arc closepath } bind def
/C3 { BL [] 0 setdash 2 copy moveto
       2 copy  vpt 0 180 arc closepath fill
               vpt 0 360 arc closepath } bind def
/C4 { BL [] 0 setdash 2 copy moveto
       2 copy  vpt 180 270 arc closepath fill
               vpt 0 360 arc closepath } bind def
/C5 { BL [] 0 setdash 2 copy moveto
       2 copy  vpt 0 90 arc
       2 copy moveto
       2 copy  vpt 180 270 arc closepath fill
               vpt 0 360 arc } bind def
/C6 { BL [] 0 setdash 2 copy moveto
      2 copy  vpt 90 270 arc closepath fill
              vpt 0 360 arc closepath } bind def
/C7 { BL [] 0 setdash 2 copy moveto
      2 copy  vpt 0 270 arc closepath fill
              vpt 0 360 arc closepath } bind def
/C8 { BL [] 0 setdash 2 copy moveto
      2 copy vpt 270 360 arc closepath fill
              vpt 0 360 arc closepath } bind def
/C9 { BL [] 0 setdash 2 copy moveto
      2 copy  vpt 270 450 arc closepath fill
              vpt 0 360 arc closepath } bind def
/C10 { BL [] 0 setdash 2 copy 2 copy moveto vpt 270 360 arc closepath fill
       2 copy moveto
       2 copy vpt 90 180 arc closepath fill
               vpt 0 360 arc closepath } bind def
/C11 { BL [] 0 setdash 2 copy moveto
       2 copy  vpt 0 180 arc closepath fill
       2 copy moveto
       2 copy  vpt 270 360 arc closepath fill
               vpt 0 360 arc closepath } bind def
/C12 { BL [] 0 setdash 2 copy moveto
       2 copy  vpt 180 360 arc closepath fill
               vpt 0 360 arc closepath } bind def
/C13 { BL [] 0 setdash  2 copy moveto
       2 copy  vpt 0 90 arc closepath fill
       2 copy moveto
       2 copy  vpt 180 360 arc closepath fill
               vpt 0 360 arc closepath } bind def
/C14 { BL [] 0 setdash 2 copy moveto
       2 copy  vpt 90 360 arc closepath fill
               vpt 0 360 arc } bind def
/C15 { BL [] 0 setdash 2 copy vpt 0 360 arc closepath fill
               vpt 0 360 arc closepath } bind def
/Rec   { newpath 4 2 roll moveto 1 index 0 rlineto 0 exch rlineto
       neg 0 rlineto closepath } bind def
/Square { dup Rec } bind def
/Bsquare { vpt sub exch vpt sub exch vpt2 Square } bind def
/S0 { BL [] 0 setdash 2 copy moveto 0 vpt rlineto BL Bsquare } bind def
/S1 { BL [] 0 setdash 2 copy vpt Square fill Bsquare } bind def
/S2 { BL [] 0 setdash 2 copy exch vpt sub exch vpt Square fill Bsquare } bind def
/S3 { BL [] 0 setdash 2 copy exch vpt sub exch vpt2 vpt Rec fill Bsquare } bind def
/S4 { BL [] 0 setdash 2 copy exch vpt sub exch vpt sub vpt Square fill Bsquare } bind def
/S5 { BL [] 0 setdash 2 copy 2 copy vpt Square fill
       exch vpt sub exch vpt sub vpt Square fill Bsquare } bind def
/S6 { BL [] 0 setdash 2 copy exch vpt sub exch vpt sub vpt vpt2 Rec fill Bsquare } bind def
/S7 { BL [] 0 setdash 2 copy exch vpt sub exch vpt sub vpt vpt2 Rec fill
       2 copy vpt Square fill
       Bsquare } bind def
/S8 { BL [] 0 setdash 2 copy vpt sub vpt Square fill Bsquare } bind def
/S9 { BL [] 0 setdash 2 copy vpt sub vpt vpt2 Rec fill Bsquare } bind def
/S10 { BL [] 0 setdash 2 copy vpt sub vpt Square fill 2 copy exch vpt sub exch vpt Square fill
       Bsquare } bind def
/S11 { BL [] 0 setdash 2 copy vpt sub vpt Square fill 2 copy exch vpt sub exch vpt2 vpt Rec fill
       Bsquare } bind def
/S12 { BL [] 0 setdash 2 copy exch vpt sub exch vpt sub vpt2 vpt Rec fill Bsquare } bind def
/S13 { BL [] 0 setdash 2 copy exch vpt sub exch vpt sub vpt2 vpt Rec fill
       2 copy vpt Square fill Bsquare } bind def
/S14 { BL [] 0 setdash 2 copy exch vpt sub exch vpt sub vpt2 vpt Rec fill
       2 copy exch vpt sub exch vpt Square fill Bsquare } bind def
/S15 { BL [] 0 setdash 2 copy Bsquare fill Bsquare } bind def
/D0 { gsave translate 45 rotate 0 0 S0 stroke grestore } bind def
/D1 { gsave translate 45 rotate 0 0 S1 stroke grestore } bind def
/D2 { gsave translate 45 rotate 0 0 S2 stroke grestore } bind def
/D3 { gsave translate 45 rotate 0 0 S3 stroke grestore } bind def
/D4 { gsave translate 45 rotate 0 0 S4 stroke grestore } bind def
/D5 { gsave translate 45 rotate 0 0 S5 stroke grestore } bind def
/D6 { gsave translate 45 rotate 0 0 S6 stroke grestore } bind def
/D7 { gsave translate 45 rotate 0 0 S7 stroke grestore } bind def
/D8 { gsave translate 45 rotate 0 0 S8 stroke grestore } bind def
/D9 { gsave translate 45 rotate 0 0 S9 stroke grestore } bind def
/D10 { gsave translate 45 rotate 0 0 S10 stroke grestore } bind def
/D11 { gsave translate 45 rotate 0 0 S11 stroke grestore } bind def
/D12 { gsave translate 45 rotate 0 0 S12 stroke grestore } bind def
/D13 { gsave translate 45 rotate 0 0 S13 stroke grestore } bind def
/D14 { gsave translate 45 rotate 0 0 S14 stroke grestore } bind def
/D15 { gsave translate 45 rotate 0 0 S15 stroke grestore } bind def
/DiaE { stroke [] 0 setdash vpt add M
  hpt neg vpt neg V hpt vpt neg V
  hpt vpt V hpt neg vpt V closepath stroke } def
/BoxE { stroke [] 0 setdash exch hpt sub exch vpt add M
  0 vpt2 neg V hpt2 0 V 0 vpt2 V
  hpt2 neg 0 V closepath stroke } def
/TriUE { stroke [] 0 setdash vpt 1.12 mul add M
  hpt neg vpt -1.62 mul V
  hpt 2 mul 0 V
  hpt neg vpt 1.62 mul V closepath stroke } def
/TriDE { stroke [] 0 setdash vpt 1.12 mul sub M
  hpt neg vpt 1.62 mul V
  hpt 2 mul 0 V
  hpt neg vpt -1.62 mul V closepath stroke } def
/PentE { stroke [] 0 setdash gsave
  translate 0 hpt M 4 {72 rotate 0 hpt L} repeat
  closepath stroke grestore } def
/CircE { stroke [] 0 setdash 
  hpt 0 360 arc stroke } def
/Opaque { gsave closepath 1 setgray fill grestore 0 setgray closepath } def
/DiaW { stroke [] 0 setdash vpt add M
  hpt neg vpt neg V hpt vpt neg V
  hpt vpt V hpt neg vpt V Opaque stroke } def
/BoxW { stroke [] 0 setdash exch hpt sub exch vpt add M
  0 vpt2 neg V hpt2 0 V 0 vpt2 V
  hpt2 neg 0 V Opaque stroke } def
/TriUW { stroke [] 0 setdash vpt 1.12 mul add M
  hpt neg vpt -1.62 mul V
  hpt 2 mul 0 V
  hpt neg vpt 1.62 mul V Opaque stroke } def
/TriDW { stroke [] 0 setdash vpt 1.12 mul sub M
  hpt neg vpt 1.62 mul V
  hpt 2 mul 0 V
  hpt neg vpt -1.62 mul V Opaque stroke } def
/PentW { stroke [] 0 setdash gsave
  translate 0 hpt M 4 {72 rotate 0 hpt L} repeat
  Opaque stroke grestore } def
/CircW { stroke [] 0 setdash 
  hpt 0 360 arc Opaque stroke } def
/BoxFill { gsave Rec 1 setgray fill grestore } def
end
}}%
\GNUPLOTpicture(2520,1512)
{\GNUPLOTspecial{"
gnudict begin
gsave
0 0 translate
0.100 0.100 scale
0 setgray
newpath
1.000 UL
LTb
200 300 M
0 63 V
0 1049 R
0 -63 V
923 300 M
0 63 V
0 1049 R
0 -63 V
1647 300 M
0 63 V
0 1049 R
0 -63 V
2370 300 M
0 63 V
0 1049 R
0 -63 V
1.000 UL
LTb
200 300 M
2170 0 V
0 1112 V
-2170 0 V
200 300 L
1.000 UL
LT0
2007 1299 M
263 0 V
243 310 M
87 9 V
44 7 V
43 6 V
43 6 V
44 6 V
43 6 V
44 6 V
43 6 V
43 5 V
44 5 V
43 5 V
44 5 V
43 5 V
43 4 V
44 5 V
43 4 V
44 4 V
43 3 V
43 3 V
44 4 V
43 3 V
44 2 V
43 3 V
43 2 V
44 3 V
43 1 V
44 2 V
43 2 V
43 1 V
44 2 V
43 1 V
44 1 V
43 1 V
43 1 V
44 0 V
43 1 V
44 0 V
43 1 V
43 0 V
44 0 V
43 0 V
44 0 V
43 0 V
43 0 V
44 0 V
43 -1 V
44 0 V
43 0 V
1.000 UL
LT1
2007 1199 M
263 0 V
243 332 M
87 28 V
44 21 V
43 19 V
43 18 V
44 17 V
43 16 V
44 15 V
43 13 V
43 13 V
44 11 V
43 10 V
44 9 V
43 8 V
43 7 V
44 5 V
43 5 V
44 5 V
43 3 V
43 2 V
44 2 V
43 2 V
44 0 V
43 1 V
43 0 V
44 -1 V
43 -1 V
44 -1 V
43 -1 V
43 -2 V
44 -1 V
43 -2 V
44 -3 V
43 -2 V
43 -2 V
44 -3 V
43 -2 V
44 -3 V
43 -2 V
43 -3 V
44 -2 V
43 -3 V
44 -3 V
43 -2 V
43 -3 V
44 -2 V
43 -3 V
44 -3 V
43 -2 V
1.000 UL
LT2
2007 1099 M
263 0 V
243 492 M
87 147 V
44 95 V
43 72 V
43 57 V
44 42 V
43 29 V
44 16 V
43 8 V
43 1 V
44 -4 V
43 -9 V
44 -11 V
43 -13 V
43 -15 V
44 -15 V
43 -16 V
44 -15 V
43 -15 V
43 -16 V
44 -15 V
43 -14 V
44 -14 V
43 -14 V
43 -13 V
44 -12 V
43 -12 V
44 -12 V
43 -11 V
43 -10 V
44 -10 V
43 -10 V
44 -9 V
43 -9 V
43 -9 V
44 -8 V
43 -8 V
44 -7 V
43 -8 V
43 -7 V
44 -6 V
43 -7 V
44 -6 V
43 -6 V
43 -6 V
44 -6 V
43 -5 V
44 -5 V
43 -5 V
1.000 UL
LT3
2007 999 M
263 0 V
767 1412 M
41 -76 V
43 -69 V
43 -59 V
44 -55 V
43 -48 V
44 -42 V
43 -38 V
43 -35 V
44 -32 V
43 -28 V
44 -26 V
43 -25 V
43 -23 V
44 -20 V
43 -19 V
44 -18 V
43 -17 V
43 -15 V
44 -15 V
43 -14 V
44 -13 V
43 -12 V
43 -11 V
44 -11 V
43 -11 V
44 -10 V
43 -9 V
43 -9 V
44 -8 V
43 -8 V
44 -8 V
43 -7 V
43 -7 V
44 -7 V
43 -7 V
44 -6 V
43 -6 V
stroke
grestore
end
showpage
}}%
\put(1957,999){\rjust{U=1.43}}%
\put(1957,1099){\rjust{U=1.41}}%
\put(1957,1199){\rjust{U=1.38}}%
\put(1957,1299){\rjust{U=1.34}}%
\put(1285,50){\cjust{$\omega / t$}}%
\put(100,856){%
\special{ps: gsave currentpoint currentpoint translate
270 rotate neg exch neg exch translate}%
\cstack{$\hbox{Im}\; \chi_q(\omega)$}%
\special{ps: currentpoint grestore moveto}%
}%
\put(2370,200){\cjust{0.15}}%
\put(1647,200){\cjust{0.1}}%
\put(923,200){\cjust{0.05}}%
\put(200,200){\cjust{0}}%
\endGNUPLOTpicture
\endgroup
 